# High-pressure synthesis of boron-rich chalcogenides B$_{12}$S and B$_{12}$Se

Kirill A. Cherednichenko,[1,2] Vladimir A. Mukhanov,[1] Aleksandr Kalinko,[3,4] and Vladimir L. Solozhenko [1,*]

[1] LSPM–CNRS, Université Sorbonne Paris Nord, Villetaneuse, 93430, France

[2] Department of Physical and Colloid Chemistry, Gubkin University, Moscow, 119991, Russia

[3] Photon Science – Deutsches Elektronen-Synchrotron (DESY), 22607 Hamburg, Germany

[4] Institute of Solid State Physics, University of Latvia, Riga, LV-1063, Latvia

* vladimir.solozhenko@univ-paris13.fr

ABSTRACT

Two boron-rich chalcogenides B$_{12}$S and B$_{12}$Se isostructural to α-rhombohedral boron were synthesized by chemical reaction of the elements at high-pressure – high-temperature conditions. The crystal structures and stoichiometries of both compounds were confirmed by Rietveld refinement and elemental analysis. The experimental Raman spectra of B$_{12}$S and B$_{12}$Se were investigated for the first time. All observed Raman bands have been attributed to the theoretically calculated phonon modes, and the mode assignment has been performed.

## INTRODUCTION

Boron-rich compounds isostructural to the α-rhombohedral boron (α-B$_{12}$) have become the subject of extensive theoretical [1-11] and experimental [12-23] studies due to their unusual properties and potential technical applications.[14,18,24] One of the most attractive features of boron-rich compounds is their outstanding mechanical properties. For instance, the reported hardness values of boron suboxide (B$_{12}$O$_2$) varies from 24 to 45 GPa.[25-31] Thus, boron suboxide is believed to be the hardest known oxide. Numerous works have been devoted to the investigation of its phase stability,[3,32] compressibility,[32-34] phonon [35-37] and thermal [34,38] properties, etc. However, other boron-rich chalcogenides, unlike boron suboxide, are still poorly studied.

Up to now, there have been only a few reports on the synthesis of boron subsulfide (B$_{12}$S$_x$) [39-41] and boron subselenide (B$_{12}$Se$_x$).[42] In all these studies boron chalcogenides were synthesized by chemical reaction between elemental boron and sulfur/selenium in graphite or tantalum crucibles at high temperatures (1200-1600°C) in Ar atmosphere. The chalcogen content "$x$" varies from 0.9 to 1.3 in $B_{12}S_x$, and from 0.9 to 1.1 in $B_{12}Se_x$. This is not surprising taking into account relatively low boiling temperatures of elemental sulfur (718 K) and selenium (958 K). Thus, the synthesis conditions (e.g. starting reagents ratios, maximum temperature, heating time, etc.) influence significantly the final stoichiometry of the products.

The loss of elemental sulfur and selenium during the high-temperature synthesis of boron chalcogenides can be prevented by applying high pressure. For instance, the high-pressure – high-temperature (HP-HT) synthesis provided a reproducible stoichiometry of many boron-rich compounds, such as $B_{12}O_2$,[21] $B_{13}N_2$,[17] $B_{50}N_2$,[43] and $B_{12}As_2$.[44] Thus, there are strong grounds to believe that HP-HT synthesis can ensure the reproducible chemical composition of boron-rich chalcogenides as well.

In the present work, we performed HP-HT synthesis of $B_{12}S$ and $B_{12}Se$. The crystal structures and chemical compositions of both boron-rich chalcogenides were confirmed by Rietveld refinement of the powder X-ray diffraction data and elemental analysis. The Raman spectra of both compounds have been experimentally observed for the first time. Based on *ab initio* calculations, we assigned all observed Raman bands.

EXPERIMENTAL

Polycrystalline samples of $B_{12}S$ and $B_{12}Se$ were synthesized at 6 GPa and 2500 K by reaction of elemental boron (Grade I ABCR) with sulfur/selenium (both Alfa Aesar, 99.5%) powders mixed in 15:1/18:1 molar ratios. The compacted mixtures of starting reagents were loaded in hBN capsules (to isolate the reaction mixture from graphite heater) and placed in the high-temperature assembly of a toroid-type high-pressure apparatus.[45] After reaching the required pressure, the reaction mixtures were heated at 2500 K for 3 min, then gradually cooled down to 1500 K during 10 min and quenched. In order to remove unreacted boron, the recovered samples were treated with 3N nitric acid (ACS, Alfa Aesar) for 20 min at 370 K, washed by deionized water and dried at 400 K.

The structure of as-synthesized compounds was studied by angle-dispersive powder X-ray diffraction at Swiss-Norwegian Beamline BM01, ESRF.[46] The wavelength of the monochromatic beam from a bending magnet was set to 0.6866 Å. Diffraction patterns were acquired during the 20 s in Debye-Scherrer geometry with rotating quartz-glass capillary using PILATUS 2M detector (high purity $LaB_6$ was used as an internal standard). The crystal structures of both compounds were refined using Maud software [47] (Fig. 1).

The chemical composition of $B_{12}S$ and $B_{12}Se$ powders was studied by energy-dispersive X-ray analysis using scanning electron microscopes FEI Quanta 200F with X-Max EDS system (Oxford Instruments) and Hitachi SU8010 with SDD X-Max$^n$ EDX system (Oxford Instruments).

Raman spectra were measured in the 100-1600 cm$^{-1}$ range at ambient conditions using Horiba Jobin Yvon HR800 spectrometer calibrated with single-crystal cubic Si at room temperature. In order to check for possible resonant effects and/or photoluminescence, the measurements were performed at two different excitation wavelengths: 473 nm and 633 nm. None of those phenomena have been observed for both compounds.

COMPUTATIONAL

The lattice parameters and atomic fractional coordinates of trigonal $R$-$3m$ $B_{12}S$ and $B_{12}Se$ phases were optimized using a linear combination of atomic orbitals (LCAO) method implemented in CRYSTAL17 code.[48] The starting unit cell parameters and atom coordinates were taken from the experimental data. Improved all-electron double-zeta valence basis sets augmented by one set of polarization functions (pob-DZVP-rev2)[49] were chosen for boron, sulfur and selenium atoms. The accuracy of the calculation of the bielectronic Coulomb and exchange series was controlled by the set of tolerances, which were set to $10^{-8}$, $10^{-8}$, $10^{-8}$, $10^{-8}$, and $10^{-16}$. The Monkhorst-Pack scheme[50] for an $8{\times}8{\times}8$ k-point mesh was used to integrate the Brillouin zone. Self-consistent field calculations were performed for hybrid DFT/HF WCGGA-PBE-16% functional.[51] The percentage of 16% defines the Hartree-Fock admixture in the exchange part of DFT functional. The tolerance for the total energy change was set to $10^{-10}$.

To calculate $B_{12}S$ and $B_{12}Se$ phonon frequencies the direct (frozen-phonon) method implemented in CRYSTAL17 code[53,54] was used. Raman intensities were calculated by using a coupled-perturbed Hartree–Fock/Kohn–Sham approach.[54,55] The parameters of the optimized unit cell as well as atomic coordinates are collected in Table IV. Raman spectra were constructed by using the transverse optical (TO) modes and by adopting a pseudo-Voigt functional form[53] with a full width half maximum parameter set to 1.

RESULTS AND DISCUSSION

**Crystal structures of $B_{12}S$ and $B_{12}Se$**

As typical $\alpha$-$B_{12}$-related compounds, boron subsulfide and boron subselenide have rhombohedral symmetry ($R$-$3m$ space group). The refined lattice parameters of $B_{12}S$ and $B_{12}Se$ were found to be in good agreement with the literature data (Table I). The unit cells of both compounds contain two independent boron atoms (in $18h$ Wyckoff positions) and one sulfur/selenium atom (in $6c$ Wyckoff position). All atomic coordinates and bond lengths are presented in Table II. The slightly distorted $B_{12}$-icosahedra are placed in the corners of the unit cell and on one of the unit cell main diagonals. S/Se atoms have a tetrahedral environment, including three B atoms belonging to three different boron icosahedra and one S/Se atom. According to the Rietveld refinement, the occupancies of $6c$ sites by S and Se atoms were found to be 55% and 52%, respectively. Since all boron atoms constitute $B_{12}$-icosahedra, the occupancies of $18h$ sites were fixed to 1.0 by default. The results of Rietveld refinement are in good agreement with energy-dispersive X-ray spectroscopy data: the elemental compositions of $B_{12}S_x$ and $B_{12}Se_x$ are 92.5(1) at% B and 7.5(1) at% S and 92.6(1) at% B and 7.4(1) at% Se, i.e. $x \approx 1$ for both compounds. The final reliability factors ($R_{wp}$) were converged to 2.4 for $B_{12}S$ and 0.5 for $B_{12}Se$, which indicates the excellent refinement level (Fig. 1).

The occupancies of $6c$ sites by S/Se atoms in $B_{12}S$ and $B_{12}Se$ refined in our study are close to the values obtained previously [40-42] (Table I). An attempt to place a single S/Se atom in $3b$ site (between two $6c$ sites) as it was proposed by Matkovich [39] resulted in not satisfactory profile fit convergence of the Rietveld refinement. Moreover, a replacement of the S/Se atom by B atom in $6c$ positions as it was done previously (e.g. $B_{12}Se_{2-x}B_x$) [42] led to significant refinement deteriorations and, hence, high $R_{wp}$ values. Thus, despite the partial occupancies of $6c$ sites in $B_{12}S/B_{12}Se$ unit cells, we exclude a possibility of the S/Se atoms "slide" to $3b$ Wyckoff position or their partial replacement by boron atoms.

The ~50% occupancy of $6c$ site in $B_{12}Se$ can be explained by the short Se-Se distance (2.02 Å) compared with the covalent radius of selenium (1.20 Å). In other words, $B_{12}Se$ unit cell is too small to accommodate two Se atoms. However, the same logic cannot be applied for $B_{12}S$: S-S interatomic distance is 2.23 Å, whereas the covalent sulfur radius is 1.05 Å. The nature of the partial occupancies of $6c$ sites in $B_{12}S$ remains unclear. As follows from Table I and Figure 2, the larger values of $6c$ site occupancy in boron subsulfides expectedly lead to the expansion of the lattice parameters (particularly, along the $c$-axis) and increase of the calculated densities. Unlike boron subsulfides, larger values of $6c$ site occupancy in boron subselenide surprisingly result in the lower calculated density and shrinking of the unit cell along the $c$-axis (Table I).

**Raman spectroscopy study of $B_{12}S$ and $B_{12}Se$**

Unlike recently synthesized orthorhombic boron-rich chalcogenides, $o$-$B_6S$ and $o$-$B_6Se$,[56] the Raman spectra of rhombohedral $B_{12}S$ and $B_{12}Se$ contain broad "noisy" bands in the 150-1250 cm$^{-1}$ range (Figure 3). According to the symmetry analysis, the acoustic and optic modes of boron-rich chalcogenides (42 atoms per unit cell) at $\Gamma$ point can be presented as follows:

$$\Gamma_{acoustic} = A_{2u} + E_u$$

$$\Gamma_{optic} = 5A_{1g} + 5A_{2u} + 7E_u + 7E_g$$

$5A_{2u} + 7E_u$ are IR active modes; $5A_{1g} + 7E_g$ are Raman active modes for the $D_{3d}$ point group.

The theoretically predicted Raman spectra of $B_{12}S$ and $B_{12}Se$ (at T = 0 K) are presented in Figure 3 (the positions of the theoretical Raman bands peaks are denoted by red dashed lines). The wavenumbers of the calculated Raman active phonon modes ($\omega'$) and the experimentally observed Raman bands ($\omega_0$) are collected in Table III. As it follows from Fig. 3 and Table III, the theoretical and experimental data are in good agreement (as it was also observed in our previous Raman studies [57,58]): the average difference between experimental and theoretical data is less than 3.5% for $B_{12}S$ (with a maximum value of 6.1% for 486.2 cm$^{-1}$ mode) and 1.5% for $B_{12}Se$ (with a maximum value of 6.3% for 477.9 cm$^{-1}$ mode).

Using the visualization procedure built in MOLDRAW software,[59] the theoretically predicted phonon modes were attributed to the oscillations of two main structural elements of $B_{12}X$, where X = S/Se: triangle composed of 'B1' polar boron atoms (further, $(B1)_3$ unit) and X atom linked with three different 'B2' equatorial boron atoms (further, X-$(B2)_3$ unit) of $B_{12}$-icosahedra (Fig. 4).

The low-frequency $A_{1g}$ and $E_g$ modes (323.3 cm$^{-1}$ and 355.2 cm$^{-1}$ for $B_{12}S$ and 199.6 cm$^{-1}$ and 231.5 cm$^{-1}$ for $B_{12}Se$) were referred to S/Se atoms oscillations (parallel and perpendicular to the $c$-axis) and the corresponding slight distortions in $B_{12}$-icosahedra. As it follows from Fig. 3$b$, $A_{1g}$ and $E_g$ modes have not been experimentally observed for $B_{12}Se$ in the 150-400 cm$^{-1}$ range. The most probable explanation for this phenomenon may be the low intensity of the corresponding Raman bands and the low signal-to-noise ratio of the whole Raman spectrum.

The $E_g$ modes of $B_{12}S$ (458.2 cm$^{-1}$; 731.8 cm$^{-1}$) and $B_{12}Se$ (449.6 cm$^{-1}$; 697.5 cm$^{-1}$) can be assigned to the tilting of $(B1)_3$ units and twisting oscillations of boron atoms in X-$(B2)_3$ unit leading to the tilting of the whole icosahedra around different unit cell directions. Meanwhile, the $E_g$ modes of $B_{12}S$ (754.1 cm$^{-1}$; 824.8 cm$^{-1}$; 1047.6 cm$^{-1}$; 1095.1 cm$^{-1}$) and $B_{12}Se$ (711.7 cm$^{-1}$; 804.3 cm$^{-1}$; 1020.7 cm$^{-1}$; 1062.8 cm$^{-1}$) correspond to symmetric and asymmetric stretching/twisting oscillations of B–B bonds and stretching/rocking oscillations of B–X bonds in $(B1)_3$ and X-$(B2)_3$ units, respectively. It might be roughly assumed that vibrations of $(B1)_3$ units lead basically to the distortions of the icosahedra and, thus, to distortions of the intra-icosahedral bonds, whereas X-$(B2)_3$ vibrations result in vibration of the inter-icosahedral bonds. It should also be noted that the amplitude of the X-$(B2)_3$ vibrations are more significant at the highest frequencies, while for the $(B1)_3$ unit it is the opposite.

The $A_{1g}$ modes of $B_{12}S$ (625.0 cm$^{-1}$; 761.9 cm$^{-1}$) and $B_{12}Se$ (645.3 cm$^{-1}$; 741.1 cm$^{-1}$) correspond to the symmetric "parasol" oscillations of X-$(B2)_3$ units, thus, resulting in distortions of the intra-icosahedral bonds. The $A_{1g}$ modes of $B_{12}S$ (1039.2 cm$^{-1}$) and $B_{12}Se$ (1024.1 cm$^{-1}$) refer to the symmetric stretching of B–X bonds in X-$(B2)_3$ unit. Finally, the $A_{1g}$ modes of $B_{12}S$ (1093.7 cm$^{-1}$) and $B_{12}Se$ (1038.4 cm$^{-1}$) were assigned to the symmetric stretching of $(B1)_3$ units. The two latter $A_{1g}$ modes of both boron-rich chalcogenides lead to the distortions of the inter-icosahedral bonds.

Taking into account a good agreement between theoretical and experimental data the phonon modes assignment provided above for the calculated Raman spectra of both compounds is also fear-relevant for the experimental spectra. The accuracy of the present mode assignment suffers from the noise and broad Raman bands that lead to the bands overlapping. Nevertheless, these features are rather typical for the Raman spectra of other boron-rich compounds (e.g. $B_4C$ [60] or $B_{12}O_2$ [35]). The more precise description of the phonon modes of $B_{12}S$ and $B_{12}Se$ might be done with the help of first- and second-order Raman scattering.[37]

## CONCLUSIONS

In the present work two new stoichiometric boron-rich chalcogenides, $B_{12}S$ and $B_{12}Se$, were synthesized at high pressure – high temperature conditions and studied by powder X-ray diffraction and Raman spectroscopy at ambient pressure. According to the Rietveld refinement of synchrotron X-ray diffraction data, both $B_{12}S$ and $B_{12}Se$ have rhombohedral symmetry and belong to the space group $R$-3$m$ (No 166). The observed Raman bands were assigned to the phonon modes and associated with the corresponding atomic movements.

## ORCID IDs

Vladimir L. Solozhenko  https://orcid.org/0000-0002-0881-9761

## NOTES

The authors declare no competing financial interest.

## ACKNOWLEDGMENTS

The authors thank Drs. I. Dovgaliuk and T. Chauveau for assistance with Rietveld analysis; and Drs. V. Bushlya and A. Jamali for help with EDX/SEM measurements. This work was financially supported by the European Union's Horizon 2020 Research and Innovation Program under Flintstone2020 project (grant agreement No 689279).

Table I.  Lattice parameters, S(Se) occupation of *6c* sites and X-ray densities of B$_{12}$S and B$_{12}$Se : present work (*pw*) and literature data.

| Compound | *a*, Å | *c*, Å | Occup., % | ρ, g/cm³ | Reference |
|---|---|---|---|---|---|
| B$_{12}$S | **5.8196(2)** | **11.9653(5)** | **55** | **2.34** | *pw* |
| | 5.80 | 11.90 | 50[*] | 2.33 | [39] |
| | 5.810(2) | 11.94(2) | 48.5 | — | [40] |
| | 5.8624(9) | 12.147(4) | 65 | — | |
| | 5.8379(6) | 12.036(1) | 59.9 | 2.36 | [41] |
| | 5.8307(5) | 12.028(2) | 60.9 | 2.37 | |
| | 5.8273(3) | 12.025(2) | 62 | 2.39 | |
| B$_{12}$Se | **5.9385(1)** | **11.9144(2)** | **52** | **2.90** | *pw* |
| | 5.9041(4) | 11.947(1) | 46.9[**] | 2.97 | [42] |

[*]   S atom occupies *3b* position (occupation is 100%), thus stoichiometry of compound is B$_{12}$S. Here, for the sake of comparison, one S atom was sheared between two *6c* sites, thus, giving occupation of 50%.

[**]   the stoichiometry of the compound is B$_{12}$Se$_{2-x}$B$_x$.

Table II.  The atomic structures and bond lengths of $B_{12}S$ and $B_{12}Se$.

| | Atom label (Wyckoff) | $x$ | $y$ | $z$ | $B_{iso}$, Å$^2$ | Site occupancy |
|---|---|---|---|---|---|---|
| $B_{12}S$ | S1 (*6c*) | 0.0000 | 0.0000 | 0.0934(2) | 3.1(1) | 0.55 |
| | B1 (*18h*) | 0.4419(2) | 0.5581(2) | 0.0437(2) | 3.6(1) | 1.0[f] |
| | B2 (*18h*) | 0.5004(3) | 0.4996(3) | 0.1948(2) | 3.9(1) | 1.0[f] |
| | S-S, Å | 2.230(20) | | | | |
| | B2-S, Å | 1.790(10) | | | | |
| | B1-B2, Å | 1.862(14) | | | | |
| | B1-B1, Å | 1.890(30) | | | | |
| | B2-B2, Å | 1.776(9) | | | | |
| | Atom label (Wyckoff) | $x$ | $y$ | $z$ | $B_{iso}$, Å$^2$ | Site occupancy |
| $B_{12}Se$ | Se1 (*6c*) | 0.0000 | 0.0000 | 0.08501(1) | 1.5(1) | 0.52 |
| | B1 (*18h*) | 0.4360(2) | 0.5640(2) | 0.0481(3) | 2.0(1) | 1.0[f] |
| | B2 (*18h*) | 0.4955(3) | 0.5045(3) | 0.1953(3) | 1.5(0) | 1.0[f] |
| | Se-Se, Å | 2.020(30) | | | | |
| | B2-Se, Å | 1.872(12) | | | | |
| | B1-B2, Å | 1.812(14) | | | | |
| | B1-B1, Å | 1.830(30) | | | | |
| | B2-B2, Å | 1.802(10) | | | | |

[f] the atom site occupancies were fixed to 1.0

Table III. The frequencies of the experimentally observed Raman bands at 633 nm excitation wavelength ($\omega_0$) and Raman active modes predicted by CRYSTAL17 ($\omega_t^C$), respectively. The overlapped bands groups (or too large bands) observed in experimental Raman spectra are presented by the corresponding frequency regions.

| $B_{12}S$ | | | $B_{12}Se$ | | |
|---|---|---|---|---|---|
| $\omega_0$, cm$^{-1}$ | $\omega_t^C$, cm$^{-1}$ | modes | $\omega_0$, cm$^{-1}$ | $\omega_t^C$, cm$^{-1}$ | modes |
| 319.2 | 323.3 | A1g | | 199.6 | A1g |
| 355.3 | 355.2 | Eg | — | 231.5 | Eg |
| 458.2 | 486.2 | Eg | 449.6 | 477.9 | Eg |
| 625.0 | 677.2 | A1g | 645.3 | 644.6 | A1g |
| 640-770 | 731.8 | Eg | 660-730 | 697.5 | Eg |
| | 754.1 | Eg | | 711.7 | Eg |
| | 761.9 | A1g | 741.1 | 745.7 | A1g |
| 813.2 | 824.8 | Eg | 812.1 | 804.3 | Eg |
| 944-1170 | 1039.2 | A1g | 900-1135 | 1020.7 | Eg |
| | 1047.6 | Eg | | 1024.1 | A1g |
| | 1093.7 | A1g | | 1038.4 | A1g |
| | 1095.1 | Eg | | 1062.8 | Eg |

Table IV. The lattice parameters of the optimized unit cell and atomic coordinates of $B_{12}S$ and $B_{12}Se$.

| | Atom label (Wyckoff) | $x$ | $y$ | $z$ | Site occupancy |
|---|---|---|---|---|---|
| $B_{12}S$ | S1 (*6c*) | 0.0000 | 0.0000 | 0.1186 | 1.0[f] |
| | B1 (*18h*) | 0.4366 | -0.4366 | 0.0479 | 1.0[f] |
| | B2 (*18h*) | -0.1787 | 0.1787 | 0.1400 | 1.0[f] |
| | $a$, Å | 5.8966 | | | |
| | $c$, Å | 12.1135 | | | |
| | V, Å³ | 364.76 | | | |
| | Atom label (Wyckoff) | $x$ | $y$ | $z$ | Site occupancy |
| $B_{12}Se$ | Se1 (*6c*) | 0.0000 | 0.0000 | 0.1217 | 1.0[f] |
| | B1 (*18h*) | 0.4352 | -0.4352 | 0.0467 | 1.0[f] |
| | B2 (*18h*) | -0.1836 | 0.1836 | 0.1397 | 1.0[f] |
| | $a$, Å | 6.0496 | | | |
| | $c$, Å | 12.1603 | | | |
| | V, Å³ | 385.42 | | | |

[f] all calculations were performed for site occupancies fixed to 1.0

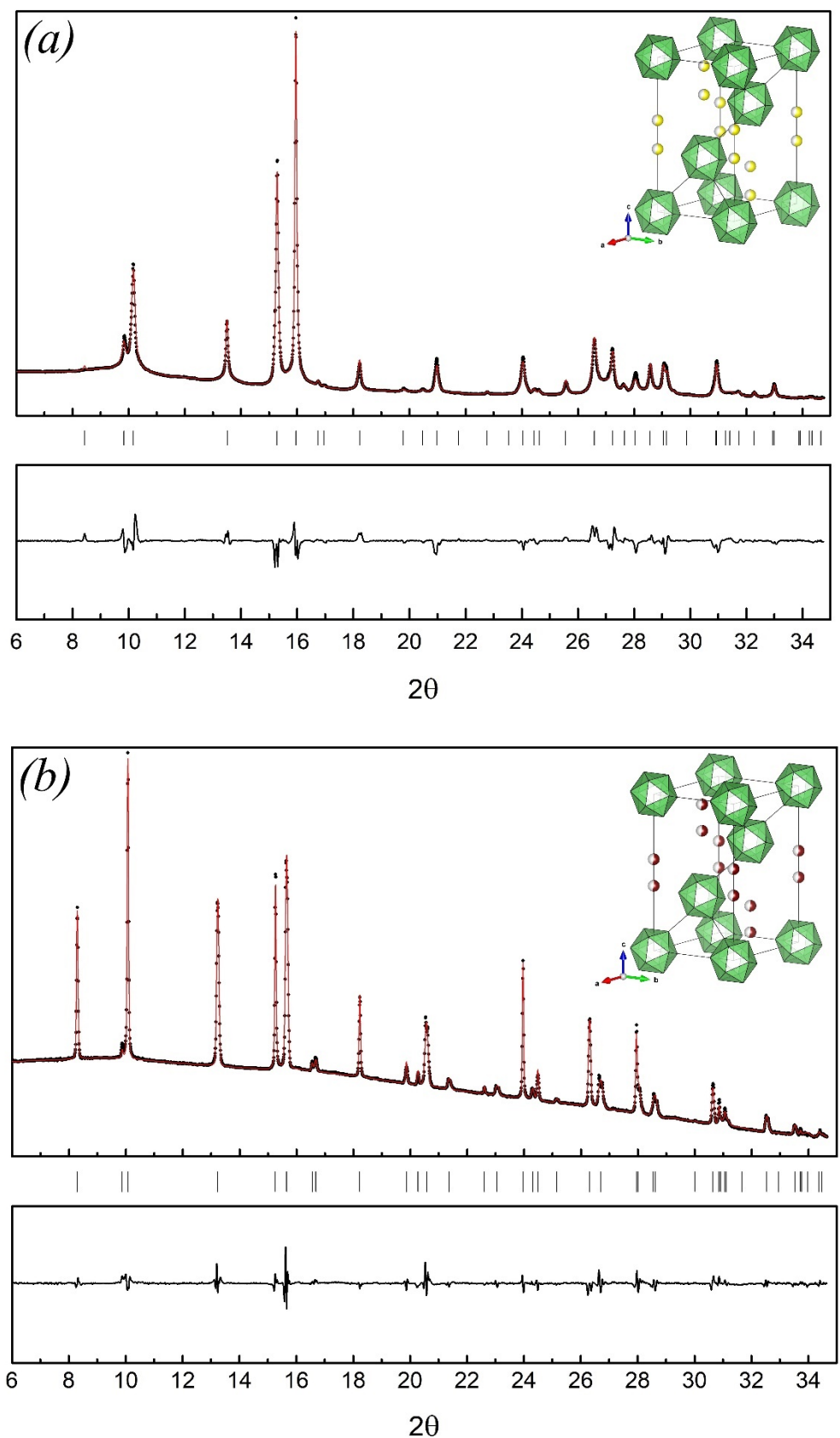

Figure 1    Rietveld full profile refinement fits of powder X-ray diffraction patterns of B$_{12}$S (*a*) and B$_{12}$Se (*b*). The insets present the unit cells of both compounds in hexagonal setting.

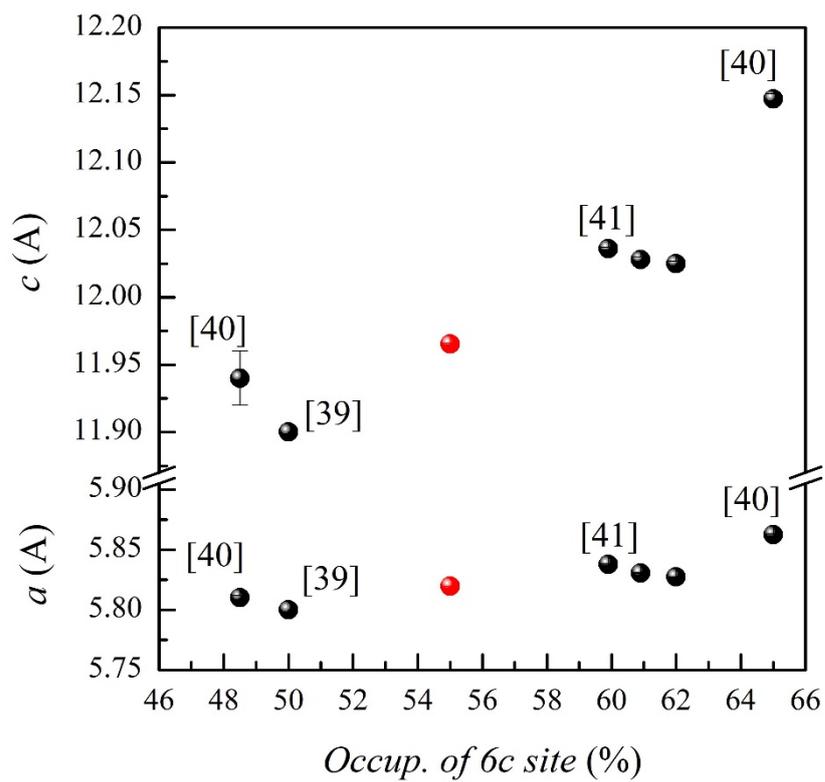

Figure 2    Comparison of B₁₂S lattice parameters *versus* occupation of *6c* site by S atom
obtained in present work (red balls) with literature data (black balls).

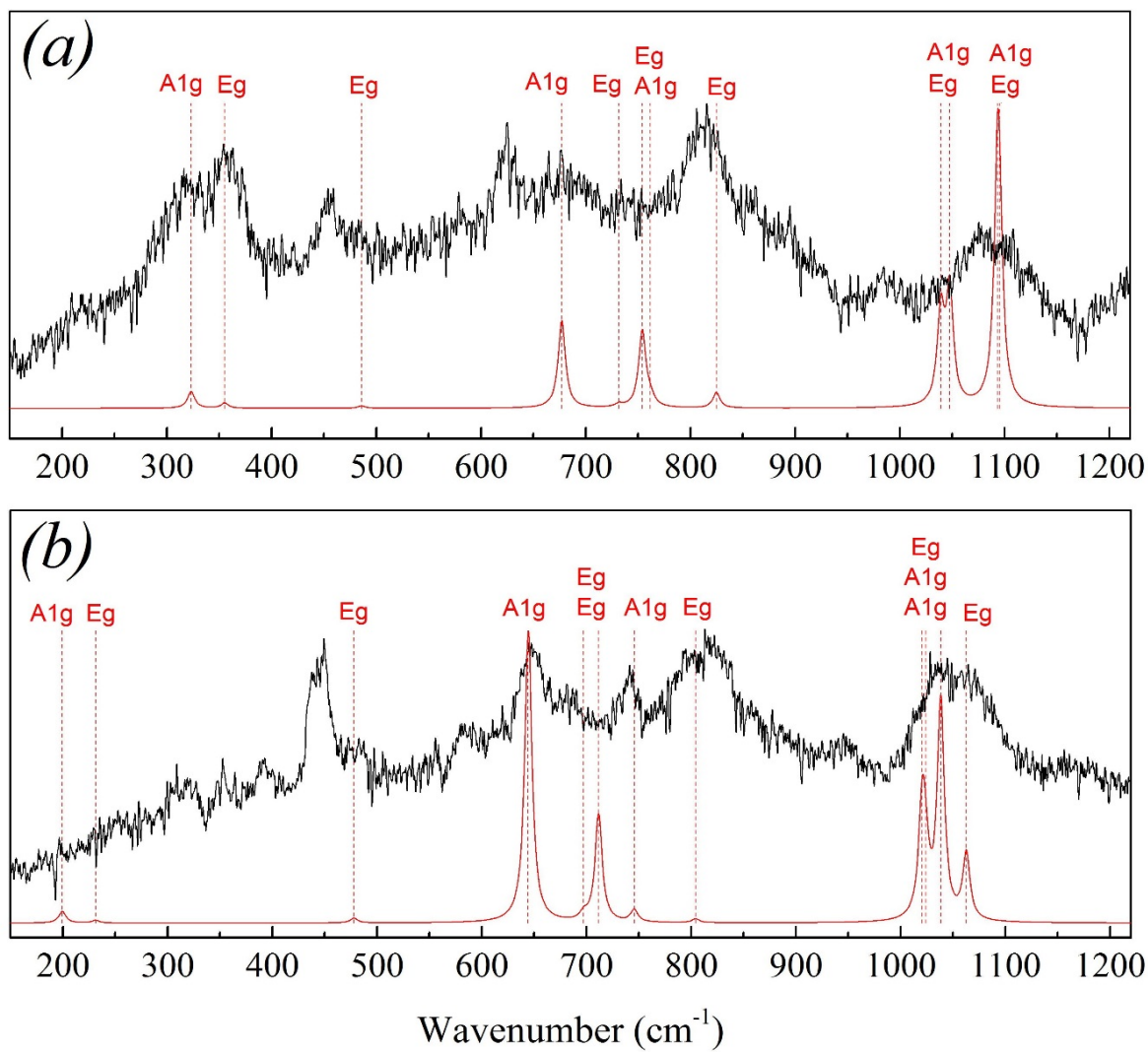

Figure 3    Experimentally observed (solid lines) and LCAO calculated (red lines) Raman bands of $B_{12}S$ (*a*) and $B_{12}Se$ (*b*) at ambient conditions. The positions of the predicted Raman peaks are traced by red dashed lines, the corresponding phonon modes are indicated.

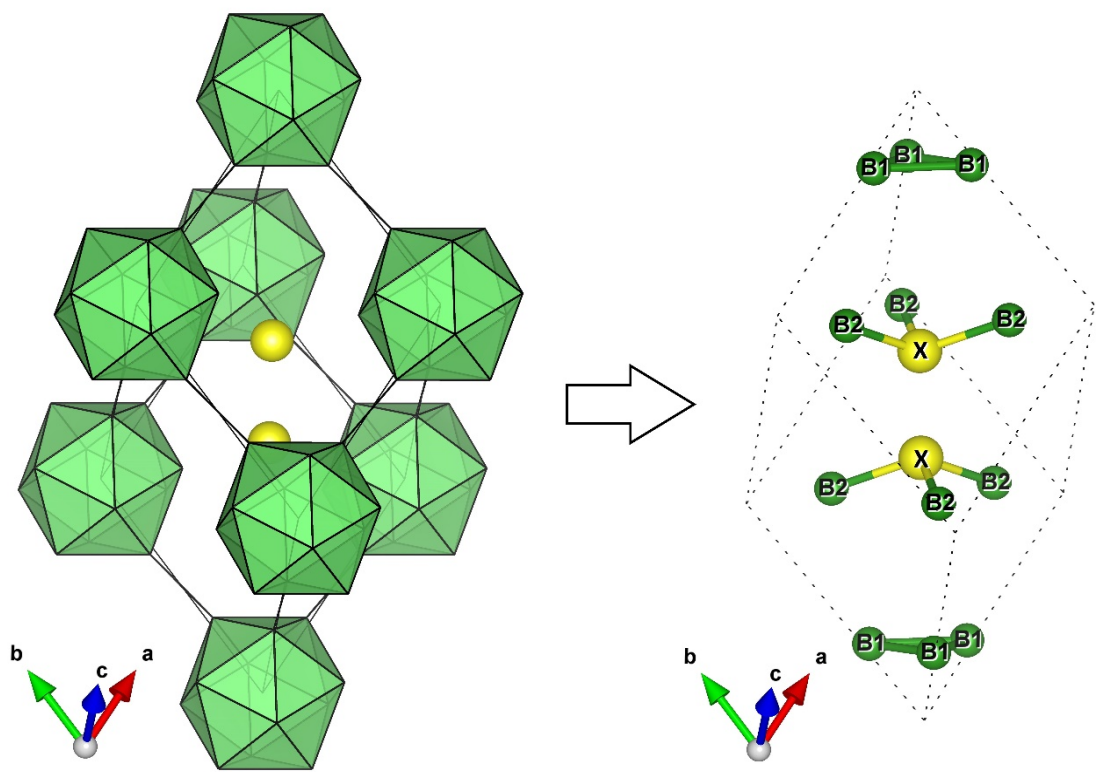

Figure 4    $B_{12}X$  (X = S, Se) unit cell in rhombohedral setting. The polar (B1) and equatorial (B2) boron atoms of $B_{12}$-icosahedra are shown by green balls, S/Se atoms are shown by yellow balls.